\documentclass[prb,aps,showpacs,floatfix,floats,nobalancelastpage,twocolumn,superscriptaddress]{revtex4}
\usepackage{amsmath, euscript}
\usepackage{amsfonts}
\usepackage{amssymb}
\usepackage{graphicx}
\usepackage{color}
\usepackage{dcolumn}
\usepackage{bm}
\usepackage{subfigure}
\usepackage{ulem}
\usepackage{CJKutf8}

\def\etal{{\it et~al}}
\def\ra{\rangle}

\def\up{\uparrow}
\def\dn{\downarrow}
\def\Hc{{\rm H.c.}}

\begin{document}

\title{Sign structures for short-range RVB states on small kagome clusters}

\author{Tiamhock Tay (\begin{CJK}{UTF8}{gbsn}郑添福\end{CJK})}
\affiliation{Department of Physics, University of California, San Diego, La Jolla, CA 92093}
\affiliation{Department of Physics, California Institute of Technology, Pasadena, CA 91125}

\author{Olexei I. Motrunich}
\affiliation{Department of Physics, California Institute of Technology, Pasadena, CA 91125}

\date{\today}



\begin{abstract}
  We extended the Schwinger boson construction to obtain wave functions that are resonating valence bond (RVB) counterparts of the degenerate coplanar classical states on the Kagome lattice. We examined all 84 of them on the 36-site cluster and found that they form a narrow energy band. On the 12-site cluster, there are only four such states and their superpositions accurately account for the second through fourth exact diagonalization (ED) states, while the ED ground state is accurately reproduced by allowing a particular two-vison insertion on top of the ${\bf q=0}$ RVB state. Thus, we have established the RVB sign structures for the low energy states on this cluster.
\end{abstract}
\maketitle


{\it Introduction.} Despite a long, enduring interest in the Heisenberg antiferromagnet on the Kagome lattice, the exact nature of the ground state has remained elusive.\cite{Ofer2006,Helton2006,Mendels2007,Olariu2007} Recent numerical studies increasingly point to a spin-disordered ground state,\cite{Yan2011, Jiang2008, Nakano2011, Lauchli2011} and the best variational state to this date is realized by the projected Dirac spin liquid.\cite{Hastings2000, Ran2007, Iqbal2010, Iqbal2011} In a series expansion study,\cite{Singh2007} Singh \etal~ found a Valence Bond Solid (VBS) state with an energy lower than the projected Dirac spin liquid. While one cannot directly compare between series expansion and variational energies, it nevertheless indicates that the true ground state might be close to a VBS instability.\cite{Evenbly2010} This is confirmed by a density matrix renormalization group (DMRG) study which found that a VBS state is indeed nearby in a generalized parameter space.\cite{Yan2011}   In a recent paper,\cite{Huh2011} Huh \etal~ investigated this possibility by considering spinless excitations of $Z_2$ spin liquids, where condensation of vortices carrying $Z_2$ gauge flux (visons) leads to a VBS phase, while gapped visons correspond to a spin liquid. In this work, we pursue a complimentary perspective by considering projected Schwinger boson (SB) wave functions with vison insertions.

{\it From classical to RVB-like states.} We construct variational wave functions by performing a Gutzwiller projection on the ground state of the following Schwinger-boson mean field Hamiltonian\cite{Sachdev1992, Auerbach, Wang2006, Tay2011}:
\begin{eqnarray}
  H_{\rm m.f.} &=& \sum_{i,j} \{A_{ij} b^\dagger_{i\downarrow} b^\dagger_{j\uparrow} + \Hc\} - \mu\sum_{i,\sigma} b^\dagger_{i\sigma} b_{i\sigma},\\
  {\bf S} &=& \frac{1}{2} \sum_{\sigma, \sigma'} b^\dagger_\sigma{\bf\mbox{\boldmath$\sigma$}}_{\sigma\sigma'}b_{\sigma'}, \\
  \vert\Psi_{\rm SB}\ra &=& \hat{\cal P}_{G} ~\exp \left\{\sum_{j,k}u_{jk} ~b^\dagger_{j\uparrow} b^\dagger_{k\downarrow} \right\} \vert 0\ra,\label{eq:RVB}\\
  u_{jk} &=& i\sum_\alpha \frac{M_{j\alpha} \lambda_\alpha (M^\dagger)_{\alpha k}}{-\mu + \sqrt{\mu^2-\lambda_\alpha^2}},\\
  \hat{A} &=& i\hat{M}\hat{\Lambda} \hat{M}^\dagger, ~~~\Lambda = {\rm diag}\{\lambda_\alpha\}.
\end{eqnarray}
Here the matrix of real-valued ``pairing amplitudes'' $A_{ij}$ is antisymmetric to produce spin-singlet states; the corresponding Hermitian matrix $-i\hat{A}$ is diagonalized to $\hat\Lambda$ by a unitary matrix $\hat{M}$. The Gutzwiller operator $\hat{\cal P}_G$ enforces the constraint $\sum_\sigma b^\dagger_\sigma b_\sigma = 1$ on each site.  In the valence bond basis, Eq.~(\ref{eq:RVB}) can be viewed as a resonating valence bond (RVB) wave function with bond amplitudes $u_{ij}$.  In the $S_z$ basis used in our variational Monte Carlo (VMC), the wave function is given by the permanent of a matrix $u_{i\in\up, j\in\dn}$.\cite{permanent}  In a recent work,\cite{Tay2011} we showed that the chemical potential $\mu$ determines the range of the RVB amplitudes $u_{ij}$ in the resulting spin liquid and performed extensive energetics studies on the original SB ansatze from Refs.~\onlinecite{Sachdev1992, Wang2006}. In this paper, we take advantage of the flexibility of the VMC setup to explore some questions accessible with more general SB constructions, including how the classical degeneracy translates to the quantum problem and how vison fluctuations may be important.

\begin{figure}
  \centering
  \begin{tabular}{cc}
    \subfigure[~${\bf q}={\bf 0}$ SB ansatz]{
      \includegraphics[trim=0 -10 0 0, scale=0.45]{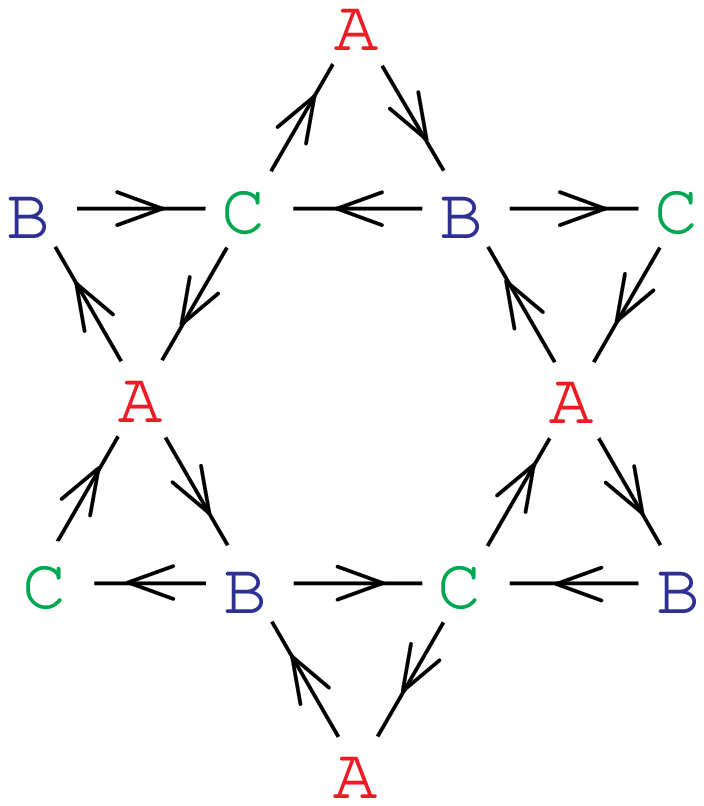}
      \label{fig:ansatz2}
    } &
    \subfigure[~$\sqrt{3}\times\sqrt{3}$ SB ansatz]{
      \includegraphics[trim=0 -10 0 0, scale=0.45]{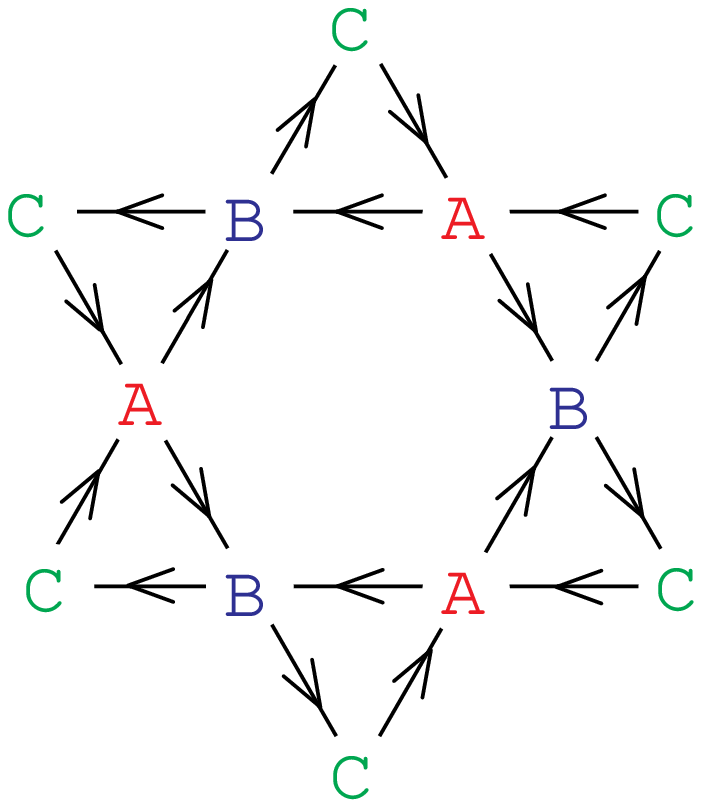}
      \label{fig:ansatz1}
    }
  \end{tabular}
  \caption{[Color online] The SB ansatze $\{A_{ij}\}$ from Ref.~\onlinecite{Sachdev1992}. $A_{ij}=1$ if an arrow points from site $i$ to $j$, and $-1$ for the opposite direction. This construction scheme readily generalizes to any ABC pattern by pointing arrows $A\rightarrow B\rightarrow C\rightarrow A$.}
  \label{fig:ansatz}
\end{figure}

We take $\mu$ and $\{A_{ij}\}$ as our variational degrees of freedom. For simplicity, from here on we restrict non-zero $A_{ij}$ to nearest-neighbor only and taking values $\pm 1$.  We generalize the symmetric ${\bf q=0}$ and $\sqrt{3}\times\sqrt{3}$ ansatze\cite{Sachdev1992,Wang2006} from our recent work\cite{Tay2011} to obtain projected wave functions that do not respect lattice symmetries.  This is motivated by the observation that the Kagome antiferromagnet has a highly degenerate manifold of classical ground states. Among these, coplanar states are believed to be most important\cite{Sachdev1992, HuseRutenberg92} and have spins pointing in one of the three coplanar directions $A$, $B$, and $C$ forming $120^\circ$ angles such that each triangle has precisely one of each label.  For each classical ground state, we obtain the corresponding ansatz as follows.\cite{Sindzingre1994} By drawing arrows between nearest-neighbor sites, pointing from $A\rightarrow B\rightarrow C\rightarrow A$, we define $A_{ij}=+1$ if an arrow points from site $i$ to $j$, and $-1$ for the opposite direction. This ``ABC'' construction scheme indeed reproduces the ${\bf q}={\bf 0}$ and $\sqrt{3}\times\sqrt{3}$ ansatze found by Sachdev (see Fig.~\ref{fig:ansatz}),\cite{Sachdev1992, Sindzingre1994} but more generally, it also admits spin liquids that break some lattice symmetries and we refer to them as ``non-symmetric''.  Note that permuting the labels $A$, $B$, and $C$ does not lead to a distinct ansatz since the direction of the arrow linking sites $i$ and $j$ remains unchanged under a cyclic permutation; for non-cyclic permutations, the directions of all arrows are reversed, but the resulting ansatz is equivalent up to a gauge choice.

\begin{figure}
  \centering
  \includegraphics[width=1.66in]{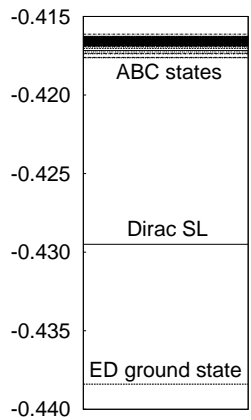}
  \caption{Variational energies per site of the 84 ``ABC'' Schwinger boson wave functions on the 36-site cluster. We include the Dirac SL and ED energies from Ref.~\onlinecite{Leung1993, Tay2011} for comparison. We set $\mu_\alpha=1.05\mu^{\rm max}_\alpha$ for each ansatz $\alpha$ in the VMC calculations. The narrow band of variational energies suggests that a projection of the Hamiltonian onto the subspace spanned by these states may allow a significant improvement in the variational ground state energy, but we did not pursue this in the present work.}
  \label{fig:ABC-energies}
\end{figure}

On the fully symmetric 36-site Kagome cluster, there are 84 distinct ansatze, and we calculate the nearest-neighbor Heisenberg exchange energy of the corresponding projected Schwinger boson wave functions using VMC. Note that for each ansatz, $\{A_{ij}\}$ are fixed at $\pm 1$ (our simplifying choice), and $\mu$ becomes the only remaining variational parameter. To allow a fair comparison between the energies of different ansatze as well as to reduce the computational cost, we fix $\mu_\alpha=1.05\mu^{\rm max}_\alpha$ for each ansatz $\alpha$, where $\mu_\alpha^{\rm max}$ is the threshold value for spinon condensation (it is negative and only $\mu < \mu^{\rm max}$ is allowed). Figure~\ref{fig:ABC-energies} shows that their energies lie within a narrow band above the exact diagonalization (ED) ground state, where the latter energy is taken from past studies.~\cite{Leung1993,Sindzingre1994,Sindzingre2009} For comparison, we also calculated Dirac SL energy for the same 36-site cluster.\cite{Ran2007, Iqbal2010, Tay2011, dirac2} The vertical scale is comparable to Fig.~4 in the ED study by Sindzingre~\etal~ which reveals many ED states populating our exhibited window.\cite{Sindzingre2009}  We remark that both the ${\bf q}={\bf 0}$ and $\sqrt{3}\times\sqrt{3}$ energies can be further lowered by varying the chemical potential $\mu$ and including further-neighbor $A_{ij}$,\cite{Sindzingre1994, Tay2011} but the lowest achieved energy within such family is about $-0.420$ per site and is only a marginal improvement on the scale of Fig.~\ref{fig:ABC-energies}. We expect roughly similar insensitivity to $\mu$ in the other ``ABC'' ansatze as well.

In a recent study,\cite{Tay2011} we included second-neighbor Heisenberg coupling $J_2$ and showed that the ${\bf q}={\bf 0}$ ansatz wins over the Dirac spin liquid for $J_2\gtrsim 0.08$. Here we also find that the $\sqrt{3}\times\sqrt{3}$ ansatz wins over the Dirac SL for $J_2\lesssim -0.05$.  The ${\bf q}={\bf 0}$ and $\sqrt{3}\times\sqrt{3}$ states have correspondingly all antiferromagnetic and all ferromagnetic second-neighbor spin correlation, while the other ``ABC'' states straddle the two limits and their energies spread out from the narrow band in Fig.~\ref{fig:ABC-energies} upon adding $J_2$.

Returning to the most challenging problem with $J_2=0$, one may reason from perturbation physics that non-vanishing off-diagonal Hamiltonian matrix elements between closely spaced energy levels might allow a substantial gain in energy. However, projecting the Heisenberg Hamiltonian into the subspace spanned by the ``ABC'' Schwinger boson wave functions is costly for the 36-site VMC calculations with permanents and is not done here.

\begin{figure}
  \centering
  \begin{tabular}{cc}
    \subfigure[~Ansatz I]{
      \includegraphics[scale=0.25]{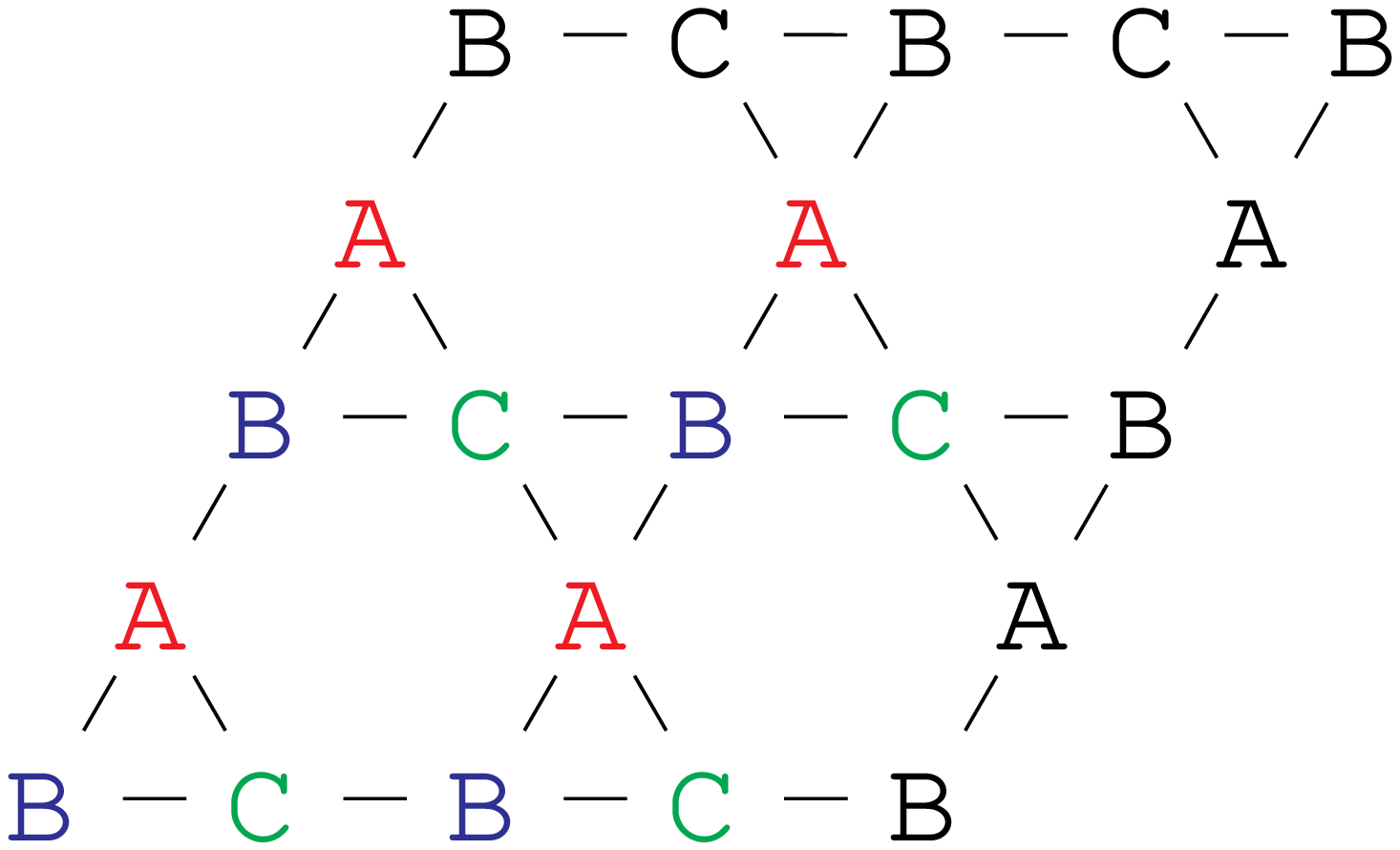}
      \label{fig:12a}
    } &
    \subfigure[~Ansatz II]{
      \includegraphics[scale=0.25]{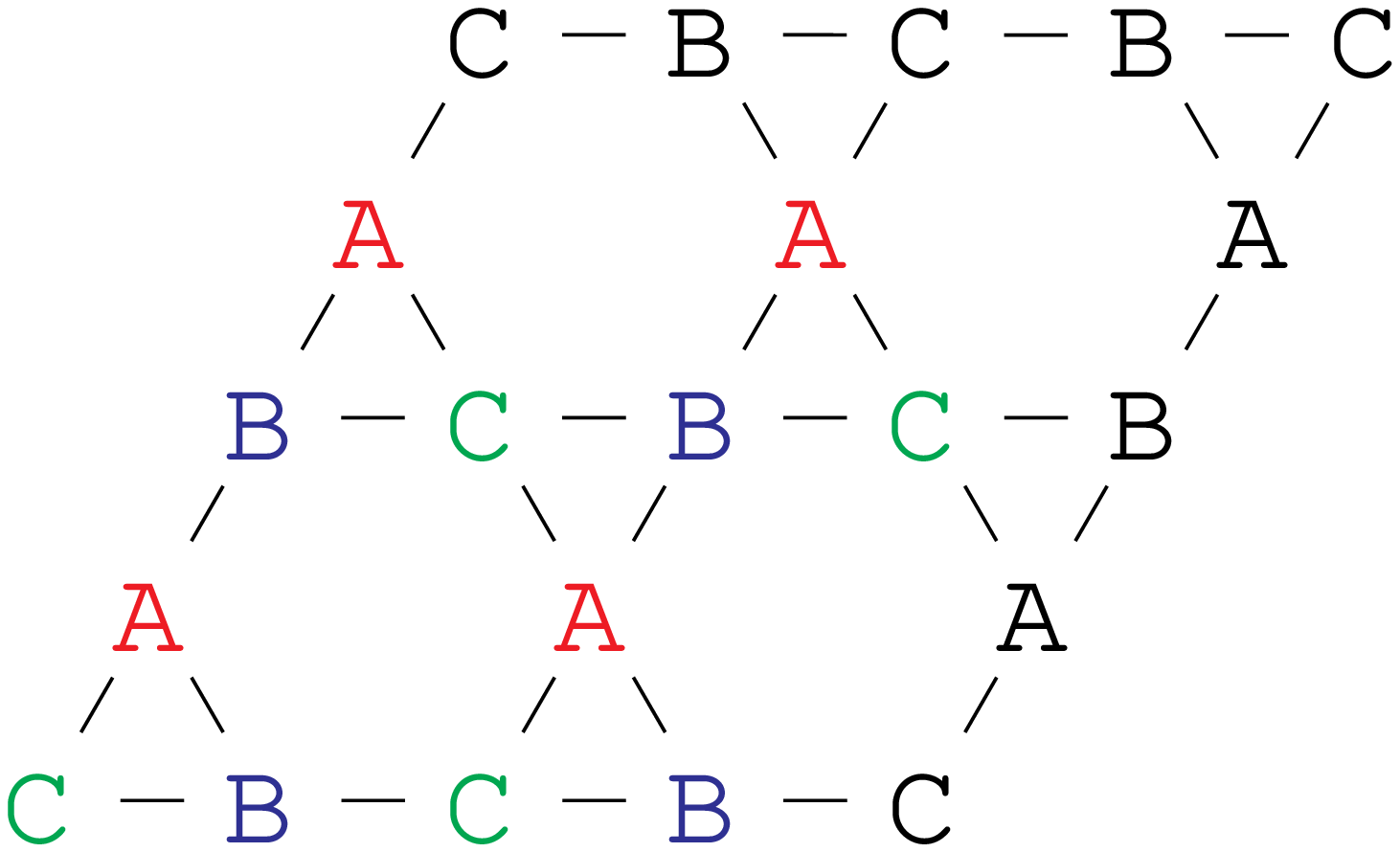}
      \label{fig:12b}
    } \\
    \subfigure[~Ansatz III]{
      \includegraphics[scale=0.25]{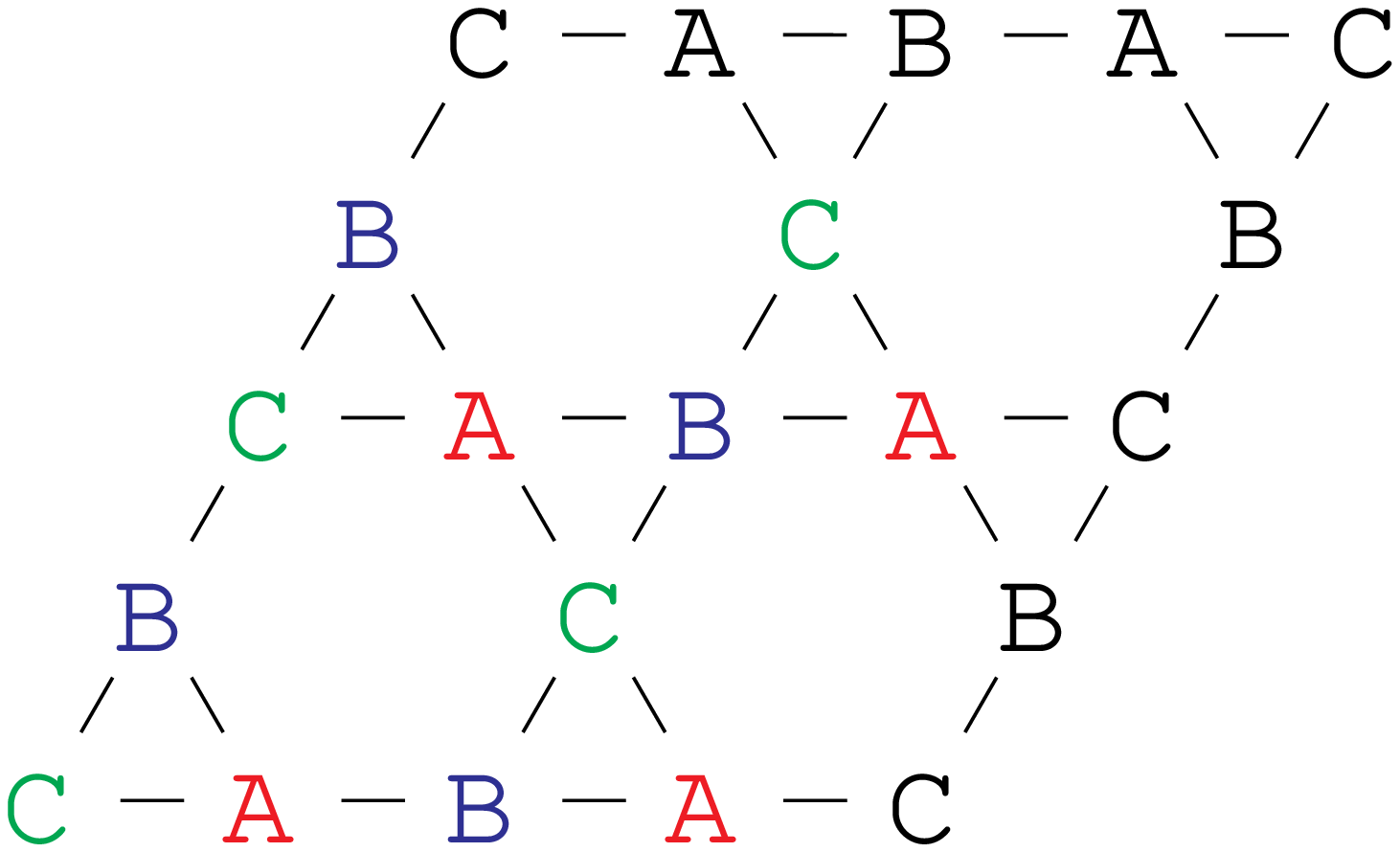}
      \label{fig:12c}
    } &
    \subfigure[~Ansatz IV]{
      \includegraphics[scale=0.25]{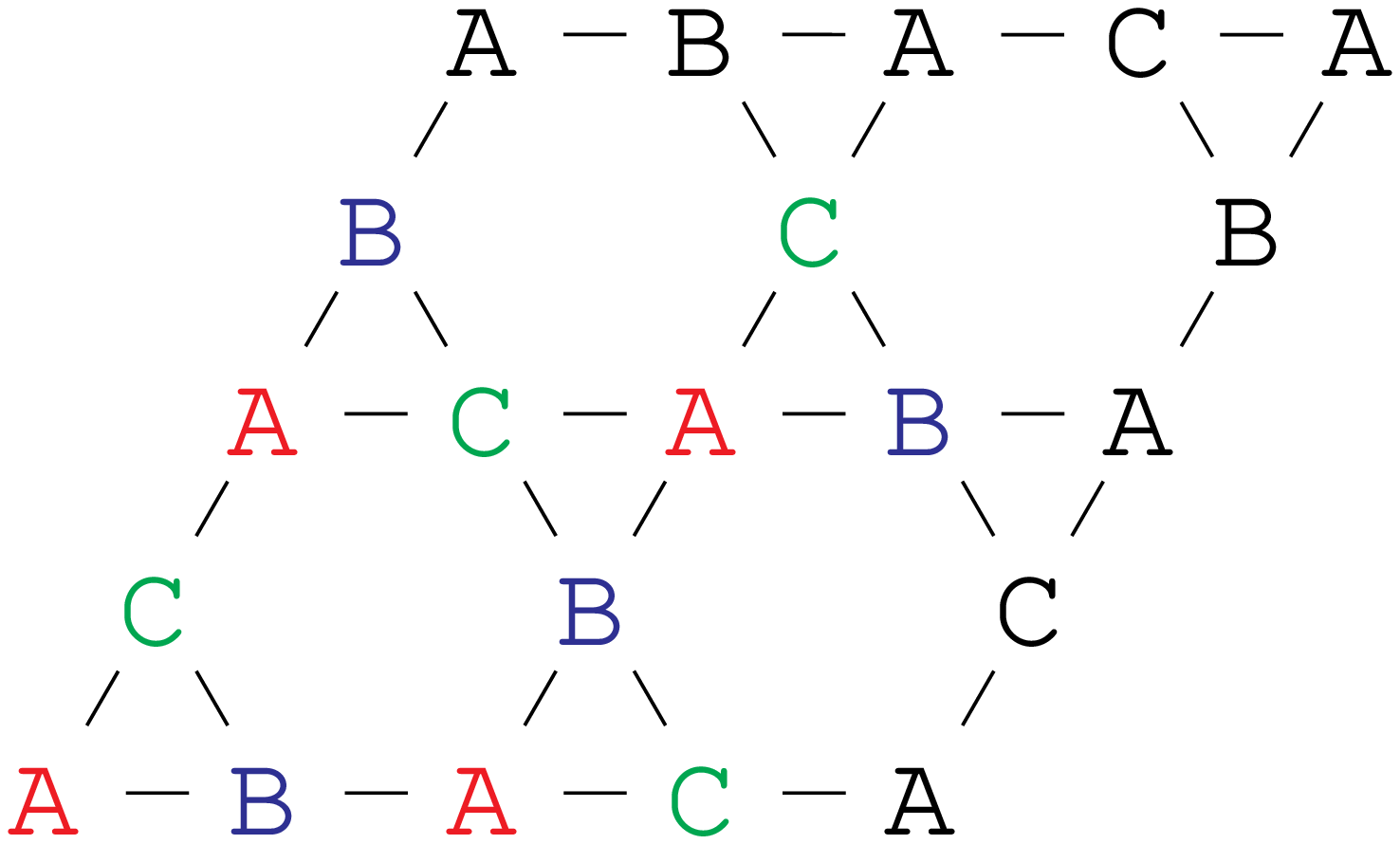}
      \label{fig:12d}
    }
  \end{tabular}
  \vskip -2mm
  \caption{[Color online] The four distinct ``ABC'' ansatze on the 12-site lattice. Colored labels mark sites on the cluster. Ansatz I can be identified with Sachdev's ${\bf q}={\bf 0}$ ansatz, while Ansatz II is obtained from Ansatz I by shifting the last row by half a horizontal lattice vector. The remaining ansatze III and IV are related to Ansatz II by $60^\circ$ rotations about the center of a hexagon.}
  \label{fig:N12}
\end{figure}

On the 12-site Kagome cluster, there are only four distinct ansatze as shown in Figs.~\ref{fig:12a}-\ref{fig:12d}. These ansatze include the ${\bf q}={\bf 0}$ ansatz, while the three remaining ansatze transform into one another under $60^\circ$ rotations. Due to the small Hilbert space, the Hamiltonian matrix elements and the wave function overlaps can be computed exactly, and we obtain the eigenvalues for the Heisenberg Hamiltonian restricted to this four-dimensional subspace.

Table~\ref{table:energies} compares the variational energies to the three lowest ED energies for the 12-site cluster. The lattice symmetries $R_{60}$, $\sigma$, $T_{\hat{e}_1}$, and $T_{\hat{e}_2}$ are the same as $C_{6}$, $\sigma$, $\vec{a}_1$ and $\vec{a}_2$ defined in Fig.~1 of Ref.~\onlinecite{Lu2011}, which refer to a $60^\circ$ rotation about a hexagon center, a reflection about the mirror passing through a hexagon center and a midpoint on its edge, and non-parallel translations that join a hexagon center to nearest neighbour hexagon centers. The optimized chemical potential $\mu_\alpha=1.01\mu^{\rm max}_\alpha$ is used for each ansatz $\alpha$. An examination of the quantum numbers shows that the best ``ABC'' state has captured the lowest ED excited state. Our analysis of the quantum numbers also shows that no linear combination of the four states has an overlap with the ED ground state. To see this, we first note that the symmetric ansatz has rotation quantum number $R_{60}=+1$. The rotational properties of the remaining ansatze only allow one to construct three eigenstates of the rotation operator with eigenvalues $+1$, $e^{\pm i2\pi/3}$. Thus, the lowest and highest states are linear combinations of the two $R_{60}=+1$ states, while the remaining states with $R_{60}=e^{\pm i2\pi/3}$ account for the two degenerate trial energies that match reasonably the degenerate ED pair in the Table. Since the rotation quantum number of the ED ground state is $-1$, there is no overlap between the ``ABC'' subspace and the ground state. The ``ABC'' states therefore realize the ED lowest excited states. 

\begin{figure}
  \centering
  \begin{tabular}{cc}
    \subfigure[~2-vison configuration I]{
      \includegraphics[trim=0 -10 0 0, scale=0.2]{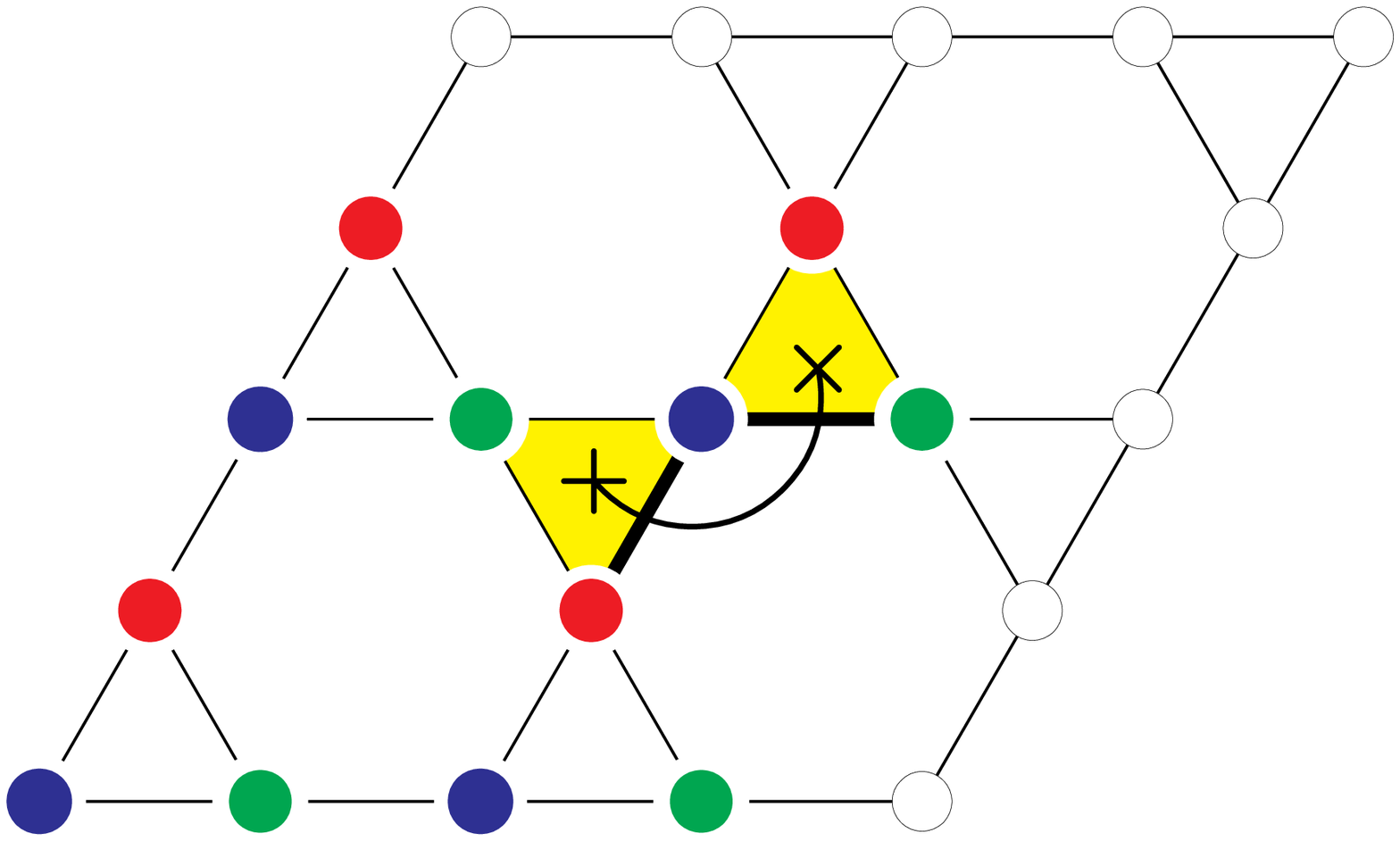}
      \label{fig:vison1}
    } &
    \subfigure[~2-vison configuration II]{
      \includegraphics[trim=0 -10 0 0, scale=0.2]{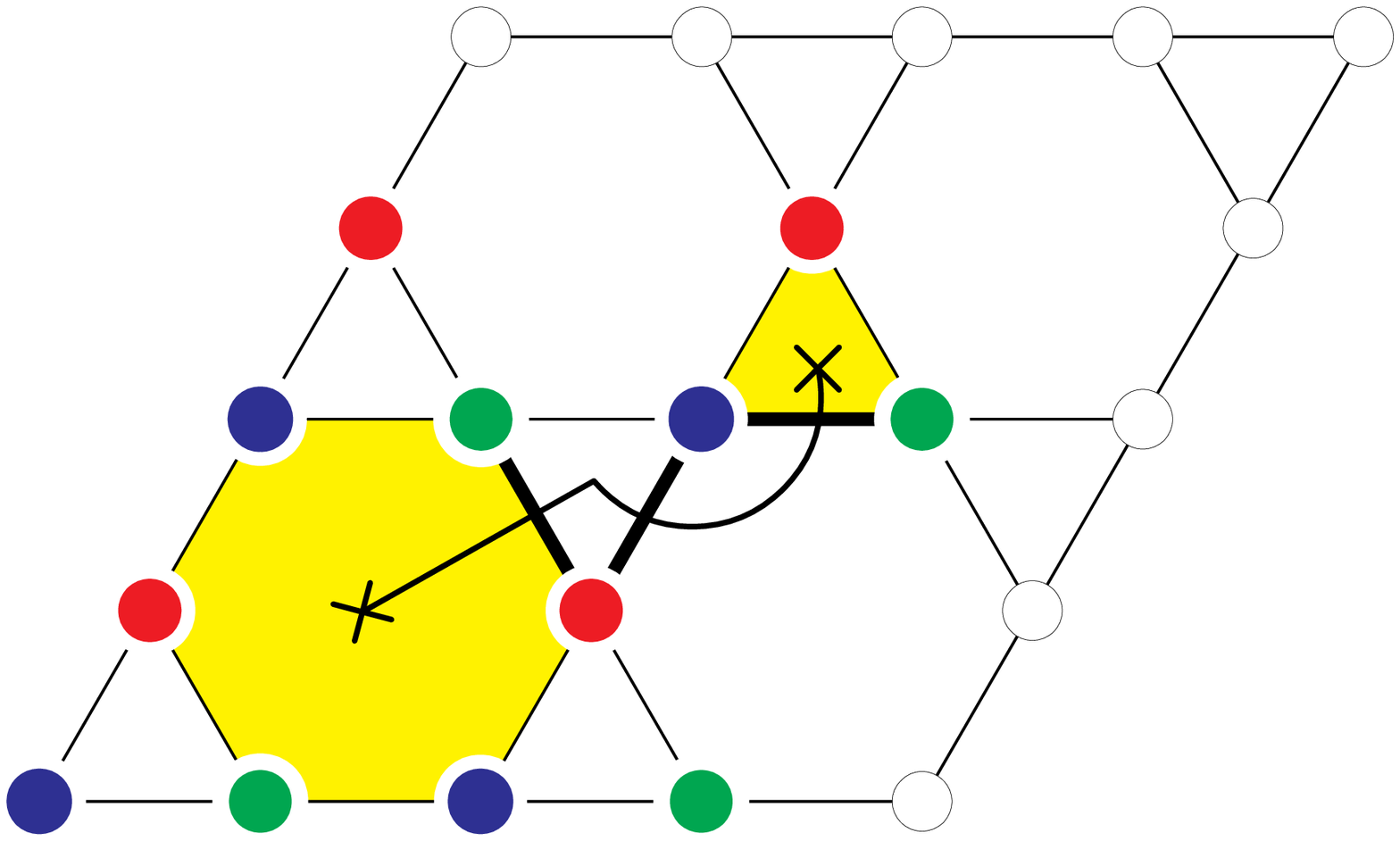}
      \label{fig:vison2}
    }
  \end{tabular}
  \caption{[Color online] The solid and open circles respectively denote sites in the 12-site cluster and equivalent sites under periodic boundary condition. The colored sites emphasize the ${\bf q}={\bf 0}$ translational symmetry of the underlying ABC ansatz. Each colored plaquette contains a vison, marked by a cross. The $A_{ij}$ on each thick link crossed by the path joining a pair of visons is multiplied by $-1$. Under translations and rotations, two-vison configuration I generates 12 distinct states, while configuration II generates 24 distinct states. Each set leads to a unique linear combination with quantum numbers $R_{60}=\sigma=-1$, $T_{\hat{e}_1}=T_{\hat{e}_2}=+1$, where $\sigma$ denotes reflection about the mirror passing through a hexagon center and a midpoint on its edge, and $T_{\hat{e}_1}$ and $T_{\hat{e}_2}$ are generators of lattice translations.}
  \label{fig:visons}
\end{figure}

{\it Schwinger boson states with vison fluctuations.} Taking a cue from the recent work of Huh~\etal,\cite{Huh2011} we consider a two-vison insertion on top of the ${\bf q}={\bf 0}$ ansatz. This is realized in our work by flipping the sign of $A_{ij}$ on the links cut by a string joining the two visons. To construct states with quantum numbers $R_{60}=\sigma=-1$, $T_{\hat{e}_1}=T_{\hat{e}_2}=+1$ on the 12-site cluster, it is sufficient to consider the ansatze generated by translations and rotations on the two-vison configurations shown in Fig.~\ref{fig:visons}. All other two-vison configurations do not lead to quantum numbers of the ED ground state. Under translations and rotations, configuration I generates 12 distinct states with visons residing on corner-sharing triangles, while configuration II gives 24 distinct states with visons residing on a hexagon and a second-neighbor triangle (to see the distinction in some cases one also needs to consider loop products of $A_{ij}$ going around the torus). For each set of ansatze, we obtain a unique (equal weight) linear combination with the same quantum numbers as the ED ground state. Thus, we have constructed trial states that contain two-vison quantum fluctuations, which apparently helps to lower the energy of the system.

\begin{table}
  \begin{tabular*}{\columnwidth}{@{\extracolsep{\fill}}cccc|ccccc}
    \hline
    ``ABC''  & 2-Vison  & SRRVB    & ED & Deg & $R_{60}$                    & $\sigma$ & $T_{\hat{e}_1}$ & $T_{\hat{e}_2}$\\
    \hline
             & -0.45313 & -0.45313 & -0.45374 & 1 & $-1$                   & $-1$     & $+1$           & $+1$ \\
             & -0.44790 &          &          & 1 & $-1$                   & $-1$     & $+1$           & $+1$ \\
    -0.44397 &          & -0.43764 & -0.44403 & 1 & $+1$                   & $+1$     & $+1$           & $+1$ \\
    -0.43384 &          & -0.42803 & -0.44152 & 2 & $e^{\pm\frac{2\pi i}{3}}$ &          & $+1$           & $+1$ \\
    -0.31649 &          &          &          & 1 & $+1$                   &          & $+1$           & $+1$ \\
    \hline
  \end{tabular*}
  \caption{Comparison between energies from ED, projected ``ABC'' wave functions, the ${\bf q}={\bf 0}$ ansatz with two-vison insertion, and short-ranged resonating valence bond (SRRVB) states from Ref.~\onlinecite{Mambrini2000}. The degeneracy (Deg) and the quantum numbers of the states are also tabulated. For the ``ABC'' states with no visons, we used optimized $\mu_\alpha=1.01\mu^{\rm max}_\alpha$.  Our ``ABC'' energy $-0.44397$ is much closer to the ED first excited level compared to $-0.43764$ obtained in Ref.~\onlinecite{Mambrini2000}, thus suggesting that the system wants longer-ranged bonds which are not realized in their nearest-neighbor dimer covers. For the two-vison states, the energies are not sensitive to $\mu_\alpha$ and we set $\mu_\alpha=-\infty$ for these cases in order to obtain better comparison with the SRRVB states; the first line corresponds to the superpositions of Fig.~\ref{fig:visons}(b) and the second line to Fig.~\ref{fig:visons}(a).
}
  \label{table:energies}
\end{table}

Table~\ref{table:energies} shows the energies of the projected Schwinger boson wave functions with two-vison insertions on top of the ${\bf q}={\bf 0}$ ansatz on the 12-site cluster. As with the ``ABC'' calculations, $\{A_{ij}\}$ are fixed at $\pm 1$ and therefore $\mu$ is the only variational parameter. We further set $\mu=-\infty$ to realize short-ranged RVB states\cite{Tay2011} (the results are essentially insensitive to $\mu$). Our second vison configuration, Fig.~\ref{fig:visons}(b), produces a variational energy that is very close to the ED ground state. The overlap with the ED ground state is 99.91\%, which is very large given that there is no variational parameter.  In the Table, we also list the energies found in an earlier study by Mambrini~\etal,\cite{Mambrini2000,Rousochatzakis2008} where the authors projected the Heisenberg Hamiltonian onto the subspace spanned by all possible short-ranged dimer-covers of the 12-site Kagome cluster. In contrast to their variational ground state which was obtained by diagonalizing the projected Hamiltonian in the large subspace, we emphasize that our construction is essentially parameter-free. Furthermore, it is difficult to understand the physics behind the sign structure of the variational ground state in the RVB basis. The simplicity of our trial zero-momentum state and its ability to reproduce their energy suggest a remarkable picture of vison dynamics hidden within the complex picture of randomly fluctuating valence bonds.

{\it Discussion.} The strong overlap between our zero-parameter projected Schwinger boson wave function with the two-vison insertion and the ED ground state on the 12-site cluster adds some support to the increasingly common view that the ground state of the Kagome antiferromagnet is close to a VBS transition.\cite{Poilblanc2011,Huh2011,Yan2011} We can contemplate a picture where the ground state contains significant local two-vison fluctuations (note that as long as individual visons are not free, the system remains a spin liquid).  It is difficult to construct workable wave functions with such local fluctuations on larger clusters, since we would expect the number of such desired pair-vison insertions to grow with the system size.  If we take the 12-site study with one pair-vison as a guide, on the 36-site cluster we would want on the order of three pair-vison insertions and correspondingly complex superpositions to produce a symmetric trial state.  

So far on the 36-site cluster, we have only considered single pair-vison insertions and constructed superpositions that produce translationally invariant states.  For a particular local pair-vison picture, we sum over lattice translations and rotations of the corresponding SB wave functions to obtain the trial state.  Such superpositions of permanents are computationally costly, but we can still report few results. 

We again take the ${\bf q}={\bf 0}$ SB state with no visons as the starting point.  Adding a static two-vison defect leads to the increase in energy; for example, for wave functions with two-vison insertions shown in Fig.~\ref{fig:visons} the trial energy became -0.403(2) per site (an increase by about 0.015 per site or 0.5 in total energy from the no-vison case, which gives some idea about the vison core energy).  When we tried translationally invariant superpositions, we found that the energy further increased for Fig.~\ref{fig:visons}(a) or remained unchanged for Fig.~\ref{fig:visons}(b), which is qualitatively different from the 12-site cluster where we observed significant lowering of energy.  We note, however, that the tried states have the mirror and rotation quantum numbers $\sigma=-1$, $R_{60}=-1$, which are different from the quantum numbers of the true ground state on the 36-site cluster.\cite{Sindzingre2009}  Thus, the defect constructions that were optimal on the 12-site cluster are not so on the 36-site cluster, indicating that finite-size effects are still important in the energetics and that we should not take the small cluster findings too literally.  The increase in energy of the zero-momentum superpositions of these defects suggests that the specific pair-vison dispersion has a band minimum at a different momentum, which we would like to examine in the future.  We have also tried several pair-vison states that have the mirror and rotation quantum numbers $\sigma=+1, R_{60}=+1$.  For example, in the case where the visons are located on a hexagon and a triangle that share a bond, the trial energy of the wave function with static defect is -0.412 per site and decreased to -0.416 per site upon forming a translationally invariant superposition.  The latter value is close to the best SB wave function energy with no defects.  It would be interesting to explore other superpositions and also think about states with several pair-vison fluctuations.

This describes our limited work so far, but it shows potential for more detailed studies of the properties of the SB spin liquid and its excitations.  For example, we can contemplate estimating vison hopping amplitudes by calculating transition matrix elements for moving a single vison, which are important ingredients in the analysis of possible vison condensation patterns in Ref.~\onlinecite{Huh2011}.

We conclude by highlighting possible extensions.  As already mentioned, it would be interesting to project the Hamiltonian into the space spanned by the 84 ``ABC'' states on the 36-site cluster.  The needed matrix element calculations are difficult in VMC but probably within reach, particularly if utilizing the expected lattice symmetries.  Using superpositions of permanents, one could also study SB wave functions on clusters with odd number of sites like the symmetric 27-site Kagome cluster.\cite{Lauchli2011}  It would also be interesting to measure energies for lattice-symmetric superpositions in the slave-fermion approaches, for example for the popular Dirac and neighboring $Z_2$ spin liquids,\cite{Hastings2000, Ran2007, Iqbal2010, Lu2011, Iqbal2011} and also allow two-vison insertions like we did for the SB construction here.  Finally, while we focused on the $J_2 = 0$ model, it would be interesting to follow the energetics upon adding small $J_2$ where we expect the SB states to perform well.\cite{Rousochatzakis2008,Messio2010,Wang2010,Tay2011}

{\it Acknowledgments}: The research is supported by the NSF through grant DMR-0907145.

\bibliography{kagome-II.bib}

\begin{thebibliography}{31}
\bibitem[{dir()}]{dirac2}
\bibinfo{note}{For the present 36-site cluster, the Dirac spin liquid that we
  used breaks lattice point group symmetries. More precisely, gauge invariant
  specification of the state also contains fluxes through loops that wind
  around the sample, and our state has such fluxes equal to 0 for two Kagome
  lattice directions and $\pi$ for the remaining direction.}
\end{thebibliography}

\end{document}